\documentstyle[12pt,a4]{article}
\newcommand{\sect}[1]{\setcounter{equation}{0}\section{#1}}

\newcommand{\beq}{\begin{equation}}

\newcommand{\eeq}{\end{equation}}
\newcommand{\beqa}{\begin{eqnarray}}
\newcommand{\eeqa}{\end{eqnarray}}
\newcommand{\beqan}{\begin{eqnarray*}}
\newcommand{\eeqan}{\end{eqnarray*}}
\newcommand{\ba}{\begin{array}}
\newcommand{\ea}{\end{array}}
\begin{document} 
\title{Post-Newtonian extension of the Newton-Cartan theory}
\author{G. Dautcourt\thanks{~E--mail: ~daut@rz.hu-berlin.de}
\\Institut f\"ur Physik, Humboldt-Universit\"at zu Berlin\\ 10115
 Berlin, Germany} 
\date{}
\maketitle
This article is dedicated to Andrzej Trautman, 
who has so clearly formulated the relations between Newton's and 
Einstein's ideas of spacetime structure and gravity.

\begin{abstract} 
The theory obtained as a singular limit of General Relativity, if 
the reciprocal velocity of light is assumed to tend to zero, is known 
to be not exactly the Newton-Cartan theory, but a slight extension 
of this theory. It involves not only a Coriolis force field, which is 
natural in this theory (although not original Newtonian), but also a scalar
field which governs the relation between Newtons time and relativistic 
proper time. Both fields are or can be reduced to harmonic functions, and 
must therefore be constants, if suitable global conditions are imposed. 
We assume this reduction of Newton-Cartan to Newton`s original theory as
starting point and ask for a consistent post-Newtonian extension and for
possible differences to usual post-Minkowskian approximation methods, as
developed, for example, by Chandrasekhar. It is shown, that both 
post-Newtonian frameworks are formally equivalent, as far as the field 
equations and the equations of motion for a hydrodynamical fluid are 
concerned.

\end{abstract}
\sect{Introduction}

Approximation methods have their place in the study of observational 
consequences of the general theory of relativity. Since Newton's  
gravity theory gives excellent results at least in the Solar system, an 
obvious and often used approach is to take this theory as a zero order 
approximation and to introduce higher order corrections if required. 
There are two different routes to define post-Newtonian approximations. 
In most cases one starts with a Newtonian potential added as some  
perturbation to a Minkowski space-time \cite{chandra},\cite{will}.
In spite of its practical success, this post-Minkowskian method has a 
certain conceptional disadvantage: Is does not really start from Newton's 
original theory. This could even be an advantage with respect to a sufficient 
accurate description of nature, since, after all, the space-time {\sl is} 
nearly Minkowskian in our surroundings. Nevertheless, it is interesting that 
an apparently quite different approach is possible. 

Since the geometrical notions used by Newton and Einstein are so different, 
one had to find a common mathematical framework for both theories. 
This was done in early papers by Friedrichs \cite{friedrichs} and 
Trautman \cite{trautman1},\cite{trautman2} and later by Ehlers 
\cite{ehlers}. 
It was shown there that Newtons theory can be cast
into a four-dimensional geometrical framework  
very similar to Einstein's theory. 

It should then be straightforward to employ this covariant formulation of 
Newton's theory, also known as Newton-Cartan theory \cite{cartan1},
\cite{cartan2}, as a starting point for an approximation method.
Indeed, an asymptotic procedure, which takes the field equations of 
General Relativity and gives as a singular limit the Friedrichs-Trautman 
formulation for large values of a parameter identified as a power of the 
reciproke velocity of light, is known \cite{daut1},\cite{daut3}. 
As noted in \cite{daut1}, for a derivation of the Newtonian theory alone it is 
not relevant whether the expansion parameter is the reciproke velocity of 
light or its square, since only diagonal terms in the metric tensor are 
involved. Off-diagonal terms which first came in as post-Newtonian corrections 
to this theory however seem to require an expansion also into odd powers of 
1/c, cf. \cite{chandra},\cite{rendall1}.

One may view the procedure geometrically as an opening of 
the light cones of spacetime, which finally become
the space-like hypersurfaces of constant Newtonian time. This picture may  
help to understand the action-at-a-distance character of the resulting 
Newtonian theory with its elliptic differential equations. Also the inverse 
singular limit $c~\to~0$ can be discussed. In this ''ultra-relativistic'' 
limit of General Relativity the light cones shrink into a congruence of 
world lines, and the limiting spacetime may be considered geometrically as a 
four-dimensional null hypersurface in a five-dimensional Riemannian space. 
In this case Einstein's field equations degenerate into {\it ordinary} 
differential equations \cite{daut2}.

The present paper is the continuation of the article      
\cite{daut3}, but now based on an expansion in powers of $1/c$
instead of $1/c^2$. To make it independent, section 2 gives a brief 
formulation of the Newton-Cartan theory as a covariant space-time formulation 
of Newtons's gravitational theory. In section 3 the Newtonian and in section 4 
the post-Newtonian limit of General Relativity is derived, together with 
a fluid description of matter. In most cases in special (Galileian) coordinates
are used, where the field equations take there simplest form. 

It was so far not clear, whether post-Newtonian corrections to the 
Newton-Cartan theory describe real phenomenae differently 
compared to usual post-Minkowskian approximation methods. 
Several colleagues suggested to the author years ago that a Newton-Cartan 
based approximation procedure would work even better.
The argument was that a full
consideration of G-dependent effects already to zero order, which we 
expect in a pure $1/c$ expansion of a theory involving both $c$ and $G$, gives 
better starting conditions for higher order terms in this expansion. However,
this argument neglects the fact that the zero order theory is nothing
else but Newtonian gravity (modulo some strange harmonic fields), 
and that a full version of Newtonian gravity is also present as a first 
approximation to the Minkowski spacetime. 
All what one eventually saves is one approximation step. 

Indead, it is shown here, that both approximation routes are formally
equivalent, although the basic spacetime concept is quite different. 

To get a better view on the physics we follow our earlier approaches 
and do not explicitly state the differentiability conditions or the class 
of functions introduced here. Topics discussed here were treated with 
more rigorous mathematical methods by Lottermoser 
\cite{lotterm1}, \cite{lotterm2} and by Rendall \cite{rendall1}.
Many conclusions stated below are correct in a strict mathematical sense, 
if weighted Sobolev spaces are used (see also \cite{bruhat}).

\sect{Newton-Cartan theory}

The Newton-Cartan theory (see, for example, \cite{cartan1},\cite{cartan2},
\cite{duval1},\cite{duval2},\cite{kuenzle1},\cite{kuenzle2},
\cite{ruede}, and for Newtonian structures arising in dimensional reductions
with a null Killing field \cite{julianicolai}) is characterized by the 
existence of a contravariant symmetric tensor field $h^{\mu\nu}$ of rank 3 and 
signature (0+++), a covariant vector field $t_\mu$, and a symmetric 
affine connection $\Gamma^{\rho}_{\mu\nu}$ on a four-dimensional manifold, 
such that 
\beq h^{\mu\nu}t_{\nu}=0,~~h^{\mu\nu}_{|\rho} =0,~~t_{\mu |\nu} =0. \eeq
The stroke denotes the covariant derivative with respect to the connection.
From the symmetry of the connection follows the existence of a scalar 
function $t(x^\mu)$, the absolute Newtonian time, such that 
$t_{\mu}=t_{,\mu}$. The connection is called Newtonian if the Trautman 
condition \cite{trautman2}
\beq h^{\lambda [ \mu}R^{\nu ]} _{(\rho\sigma)\lambda} = 0 \eeq
is satisfied for the Riemann tensor contructed from the connection. 
 (2.2) may be considered as part of the field equations. The 
remaining part is given by (we follow the sign conventions of the book by 
Misner, Thorne and Wheeler \cite{mtw}) 
\beq R_{\mu\nu} = R^\rho_{\mu\rho\nu} = 4 \pi G \rho t_{\mu}t_{\nu}, \eeq
where $\rho$ is the Newtonian matter density. For a simple perfect fluid,
one has to add the equations of motion and conservation of mass, which 
do not follow from the field equations in a pure Newtonian setting:
\beq \rho v^{\mu}_{|\nu}v^\nu = -h^{\mu\nu}p_{,\nu}, \label{mat}\eeq
\beq v^\mu\rho_{,\mu}+ \rho v^\mu_{|\mu}= 0. \label{mom}\eeq

We have already written down all basic relations of the Newton-Cartan theory.
Notice that the relations (2.1) do not fix the affine connection. This is
clearly seen in special coordinates with $x^0=t,
t_{\mu}=\delta^0_{\mu}$, determined up to transformations 
\beq t'=t+ const, x^{i'} = x^{i'}(x^i,t).\eeq
In these coordinates $h^{\mu\nu}$ reduces to a contravariant 
3-tensor $h^{ik}$ describing the intrinsic metric of the hypersurfaces 
$t=const$ of equal Newtonian time $t$. The equations (2.1) become
\beq  \Gamma^{0}_{\mu\nu} =0, \Gamma^{i}_{00} = h^{ik}\Phi_k, 
\Gamma^{i}_{0k} = h^{il}\omega_{kl} + h^{il}\dot{h}_{kl}/2, 
\Gamma^{i}_{kl} = \{^{i}_{kl}\}, \eeq
where $h_{ik}$ is the inverse metric to $h^{ik}$ and  $\{^{i}_{kl}\}$
are the Christoffel symbols to the three-dimensional metric $h_{ik}$,
a dot denotes the time derivative. The dynamical 
variables of the theory are a vector field $\Phi_k$, the acceleration field, 
and a pseudo vector field
$\omega_{kl}= -\omega_{lk}$, the Coriolis force field. The intrinsic metric 
of the hypersurfaces 
$t=const$ is determined by the spatial part of the field equations (2.3), 
which split into (a double stroke denotes covariant derivations with
$ \dot{h}_{kl\|i}$)
\beqa R_{kl}(h_{kl}) &=&0, \\
h^{kl}(\dot{h}_{li\|k}-\dot{h}_{lk\|i})-  h^{kl}\omega_{li\|k} &=&0, \\
h^{kl}\Phi_{k\|l}+ \omega_{kl}\omega^{kl} -\ddot{h}_{kl}h^{kl}/2 
-\dot{h}_{kl}\dot{h}^{kl}/4 &=& 4\pi G\rho.
\eeqa
It is seen that the Ricci tensor and hence here, in three dimensions, 
the Riemann tensor of the Newtonian hypersurfaces $t=const$ is zero, 
so $h^{ik}$ is flat and can therefore be transformed into the constant
values
$\delta_{ik}$ in suitable coordinates (called Galilean coordinates)
everywhere on the manifold. Transformations preserving the Galilean character 
of these coordinates constitute the kinematical group
\beq x'^i = q^{ik}x^k + q^i, t' = t + const\eeq
with 
\beq q^{li}q^{lk} = q^{il}q^{kl} = \delta^{ik}, q^{ik}=q^{ik}(t), 
q^i=q^i(t),\eeq
corresponding to arbitrarily accelerated and rotated Cartesian coordinate 
systems.

Compared with Newton's classical theory, the Newton-Cartan theory appears to
have more dynamical degrees of freedom. Local cosmological solutions involving 
these additional degrees are discussed in \cite{daut4} (see also 
\cite{heckmann},\cite{heckmannschuecking},\cite{ruede} for related 
cosmological questions). If Trautman's condition (2.2) is taken into 
account, which becomes \beq  \Phi_{i,k}- \Phi_{k,i}= -2 \dot{\omega}_{ik}, 
~~\omega_{ik,l}\epsilon^{ikl}=0 \eeq
in special coordinates,
it is easy to see from (2.9) and (2.13b), that the
Coriolis force field can be derived from a harmonic function $c$  with
$\omega_{ik}= \epsilon_{ikl}c_{,l}.$
Given suitable boundary conditions, this function must be a spatial constant 
(see the discussion in section 3). The Coriolis force then vanishes, and
the acceleration force derives from the gradient of a Newtonian potential
$\Phi$ (see (2.13a), which in Galilean coordinates satisfies the familiar 
potential equation
\beq \Phi_{,kk}= 4\pi G\rho.\eeq
Under these assumptions, the Newton-Cartan theory reduces to Newton's original 
theory. 

\sect{Newtonian limit of GR}
To derive the Newton-Cartan theory of the last section from the field 
equations of General Relativity, consider a family $g^{\mu\nu}(x^\mu,\eta)$ 
of contravariant metrics, which depend smoothly on a parameter $\eta$. Let 
the signature of $g^{\mu\nu}$ be $(-+++)$ for every $\eta \not= 0$, and let
$g^{\mu\nu}$ for $\eta \to 0$ tend to a limit $h^{\mu\nu}$ of rank 3 and
signature (0+++). $h^{\mu\nu}$ has just one eigenvector $t_\mu$ with zero
eigenvalue:
\beq h^{\mu\nu}t_\nu = 0. \eeq
For small $\eta$ the contravariant metric components $g^{\mu\nu}(x^\mu,\eta)$ 
may be written as a power law expansion in $\eta$:
\beq g^{\mu\nu}(x^\mu,\eta)= h^{\mu\nu} + \eta h^{\mu\nu}_{(1)}
+ \eta^2 h^{\mu\nu}_{(2)} + \eta^3 h^{\mu\nu}_{(3)} +o(\eta^4). \eeq
The corresponding covariant components of the metric tensor 
$g_{\mu\nu}(x^\mu,\eta)$ should admit an asymptotic representation 
starting with negative powers of $\eta$:
\beq g_{\mu\nu}(x^\mu,\eta)= \alpha_{\mu\nu}/\eta^2 + \beta_{\mu\nu}/\eta 
+ \gamma_{\mu\nu} +\eta\gamma_{\mu\nu}^{(1)} +\eta^2\gamma_{\mu\nu}^{(2)} 
+o(\eta^3). \eeq
From $g_{\mu\alpha}g^{\nu\alpha} =\delta^\nu_\mu$ one obtains a number of 
relations between co- and contravariant quantities:
\beqa h^{\mu\rho}\alpha_{\nu\rho} &=&0, \\  
h^{\mu\rho}\beta_{\nu\rho}+ h^{\mu\rho}_{(1)}\alpha_{\nu\rho} &=&0,\\  
h^{\mu\rho}\gamma_{\nu\rho} + h^{\mu\rho}_{(1)}\beta_{\nu\rho}
+ h^{\mu\rho}_{(2)}\alpha_{\nu\rho} &=& \delta^\mu_\nu, \\
h^{\mu\rho}\gamma_{\nu\rho}^{(1)} + h^{\mu\rho}_{(1)}\gamma_{\nu\rho} 
+h^{\mu\rho}_{(2)}\beta_{\nu\rho}+h^{\mu\rho}_{(3)}\alpha_{\nu\rho} &=& 0,\\
h^{\mu\rho}\gamma_{\nu\rho}^{(2)} +h^{\mu\rho}_{(1)}\gamma_{\nu\rho}^{(1)} 
+ h^{\mu\rho}_{(2)}\gamma_{\nu\rho}+ h^{\mu\rho}_{(3)}\beta_{\nu\rho} 
+h^{\mu\rho}_{(4)}\alpha_{\nu\rho} &=& 0. \eeqa
Equation (3.4) shows that $\alpha_{\mu\nu} = at_{\mu}t_{\nu}$, and multiplying 
(3.5) with $t_\mu$ gives $ah^{\rho\sigma}t_\rho t_\sigma =0$. Thus either 
$a=0$ or $h^{\rho\sigma}t_\rho t_\sigma =0$. With $a=0$, the covariant 
metric tensor would start with a $1/\eta$ component. This corresponds to 
the procedure considered in \cite{daut1},\cite{daut3}, where $\eta=1/c^2$. 
Here we assume $a \not= 0$ and $h^{\rho\sigma}_{(1)}t_\rho t_\sigma =0$ 
as well as $\eta=1/c$. Furthermore, in order to simplify the treatment, 
we assume $\beta_{\mu\nu}=0$ subsequently. Taking $\beta_{\mu\nu}$ different 
from zero produces non-Newtonian shadow fields, which must be eliminated 
anyway by using the field equations. 

Absorbing the factor $a$ into $t_{\mu}$, we can write
\beq \alpha_{\mu\nu} = -t_{\mu}t_{\nu}, \eeq
the minus sign is necessary for a correct signature of $g_{\mu\nu}$.
(3.5) shows that also $h^{\mu\nu}_{(1)}$ is degenerate and admits the same
eigenvector $t_{\mu}$ as $h^{\mu\nu}$. The fields describing Newton's 
approximation level are hidden in the expansion term $\gamma_{\mu\nu}$,
thus $\gamma^{(1)}_{\mu\nu}$ and $\gamma^{(2)}_{\mu\nu}$ should describe
post-Newtonian effects. As we shall see, $\gamma^{(1)}_{\mu\nu}$ contains
only shadow fields which are apparently not relevant physically. In the 
contravariant components the fields describing different approximation levels
are spread over a number of terms. For example, post-Newtonian fields
are hidden in the time-time component of $h^{\mu\nu}_{(6)}$, in the space-time 
components of $h^{\mu\nu}_{(4)}$ and in the space-space components of 
$h^{\mu\nu}_{(2)}$.

Concomitants of the metric tensor are in general singular for $\eta \to 0$.
The Christoffel symbols admit a representation
\beq
\Gamma^{\mu}_{\rho\sigma} = \Gamma^{\mu(-2)}_{\rho\sigma}/\eta^2 +
\Gamma^{\mu(-1)}_{\rho\sigma}/\eta + \Gamma^{\mu(0)}_{\rho\sigma}/\eta +
\eta \Gamma^{\mu(1)}_{\rho\sigma} + \eta^2 \Gamma^{\mu(2)}_{\rho\sigma} +
o(\eta^3) \eeq
and the Riccitensor starts with $1/\eta^4$ terms: 
\beq R_{\mu\nu}=  R^{(-4)}_{\mu\nu}/\eta^4 + R^{(-3)}_{\mu\nu}/\eta^3 +
R^{(-2)}_{\mu\nu}/\eta^2 + R^{(-1)}_{\mu\nu}/\eta  + R^{(0)}_{\mu\nu}       
+ \eta R^{(1)}_{\mu\nu} + \eta^2 R^{(2)}_{\mu\nu} + o(\eta^3) \eeq
For every $\eta$, the metric 
is constraint to satisfy the field equations, which we write with 
\beq T_{\mu\nu}^{*}= \kappa (T_{\mu\nu}-\frac{1}{2}g_{\mu\nu}T) \eeq
as
\beq R_{\mu\nu} = T_{\mu\nu}^{*}.\eeq
The behaviour of $T^*_{\mu\nu}$ for $\eta \to 0$ depends on the fields 
which constitute the matter tensor. For a perfect fluid, which will be 
assumed throughout this paper, $T_{\mu\nu}^{*}$ tends to a finite limit 
for $\eta \to 0$ (see the first Appendix). This means that all singular 
terms in the Ricci tensor expansion (3.11) must vanish separately.  

We first calculate the coefficients of the expansion (3.10), which 
are easily derived from (3.2), (3.3), using the relations (3.1), (3.9).
With the notation $t_{[\mu,\nu]} = \frac{1}{2}(t_{\mu,\nu}-t_{\nu,\mu})$, 
one obtains
\beqa \Gamma^{\mu(-2)}_{\rho\sigma} &
=& h^{\mu\lambda}(t_\rho t_{[\sigma,\lambda]}
+ t_\sigma t_{[\rho,\lambda]}), \\
\Gamma^{\mu(-1)}_{\rho\sigma} &=& 
h^{\mu\lambda}_{(1)}(t_\rho t_{[\sigma,\lambda]}
+t_\sigma t_{[\rho,\lambda]}),\\ 
\Gamma^{\mu(0)}_{\rho\sigma} &=&
\frac{1}{2}h^{\mu\lambda}(\gamma_{\lambda\rho,\sigma}
+\gamma_{\lambda\sigma,\rho}-\gamma_{\rho\sigma,\lambda})\nonumber\\ 
&& + h^{\mu\lambda}_{(2)}(t_\rho t_{[\sigma,\lambda]} 
+ t_\sigma t_{[\rho,\lambda]}  
 - \frac{1}{2}t_{\lambda}(t_{\rho,\sigma}+t_{\sigma,\rho})) \eeqa
The lowest order coefficients in the Riccitensor expansion (3.11)
come from terms which are nonlinear in the Christoffel symbols:   
\beq R^{(-4)}_{\mu\nu} = 
h^{\alpha\beta}h^{\rho\sigma}t_{[\alpha,\rho]}t_{[\beta,\sigma]}. \eeq
Putting (3.17) equal to zero is equivalent to 
\beq t_{\mu} = ft_{,\mu},\eeq
cf. \cite{daut3}. Thus the vector $t_\mu$ can be reduced to two scalar 
functions $f(x^\mu), t(x^\mu)$. 
With (3.18), $R^{(-3)}_{\mu\nu}=0$ is already satisfied, and the leading 
coefficients in the connection become
\beq \Gamma^{\mu(-2)}_{\rho\sigma} = 
f f_{,\nu} h^{\mu\nu}t_{,\rho} t_{,\sigma},
\Gamma^{\mu(-1)}_{\rho\sigma} = 
f f_{,\nu} h^{\mu\nu}_{(1)}t_{,\rho} t_{,\sigma}, \eeq
The first non-vanishing coefficients of the Riccitensor are now
$R_{\mu\nu}^{(-2)}$ and $R_{\mu\nu}^{(-1)}$. 
To extract the content of $R^{(-2)}_{\mu\nu}=0$, it is appropriate to 
choose a special coordinate system with $x^0=t$. From (3.6) one obtains 
(latin indices take the values 1,2,3)
\beq h^{0\mu}=0, h^{il}\gamma_{kl}= \delta_{k}^{i}, h_{(2)}^{00}= -1/f^2,
h_{(2)}^{0k}= \gamma_{0l}h^{kl}.  \eeq
After some simplifiations using (3.20), $R^{(-2)}_{00}=0$ takes the simple 
form of a Laplacian applied to 
the scalar function $f$ in the three-dimensional space $t=const$ 
with the intrinsic metric $\gamma_{ik}$
\beq     \Delta f = 0.\eeq
We have obtained the Newton-Cartan theory in a slightly extended form, 
involving an additional function $f$ called time function, since
the relation between Newtonian time $t$ and the relativistic proper
time $s$ is given by $ds/dt= cf(x^i,t)$. Its physical relevance is not
clear. $f$ must be a harmonic function, a completely source-free solution 
of the covariant Laplace equation. There exist many local 
solutions of this equation, for example harmonic polynomials, but one 
expects that they are generated by sources near spatial infinity. 
If suitable global conditions such as a constant boundary value at 
spatial infinity or the compactness of the hypersurfaces $t=const$ 
are imposed, one can conclude that $f$ is a constant with respect 
to spatial coordinates. Without loss of generality we may assume 
$f=1$. Then {\it all} conditions $R^{(-2)}_{\mu\nu}=0$ are satisfied, 
moreover, this also holds for the relations $R^{(-1)}_{\mu\nu}=0$. 
The Christoffel symbol expansion now starts with the finite term
\beq \Gamma^{\mu(0)}_{\rho\sigma} =
\frac{1}{2}h^{\mu\lambda}(\gamma_{\lambda\rho,\sigma}
+\gamma_{\lambda\sigma,\rho}-\gamma_{\rho\sigma,\lambda})
- h^{\mu\lambda}_{(2)}t_{,\lambda}t_{,\rho\sigma},  \eeq
which transforms as an affine connection under arbitrary coordinate 
transformations in spite of its unusual appearence. 
The Ricci tensor calculated to lowest order reduces to the Ricci tensor
build from the connection (3.22), and the field equations take with (5.3) 
the form (2.3).
We have recovered the Newton-Cartan theory, with (3.22) as 
a Newtonian connection. The relations (2.1) and (2.2) may be checked 
in a special coordinate system $x^0=t$.

\sect{Post-Newtonian extension}

Going now to the first post-Newtonian order, we have first to consider 
the expansion components 
$\Gamma^{\rho (1)}_{\mu\nu}$  of the connection. Subsequently, we use always
the special coordinate system with $t=x^0$ and $\gamma_{ik}= \delta_{ik}$. 
From (3.7) follows $h_{(1)}^{ik} = - \gamma^{(1)}_{ik}$,
and a straightforward calculation gives
\beqa \Gamma^{0(1)}_{\mu\nu} &=& 0 \\
\Gamma^{i(1)}_{00} &=& \frac{1}{2}(2\gamma^{(1)}_{i0,0} - 
\gamma^{(1)}_{00,i} \\ \Gamma^{i(1)}_{0k} &=& 
\frac{1}{2} \gamma^{(1)}_{ik,0} + \frac{1}{2} \gamma^{(1)}_{0i,k} -
\gamma^{(1)}_{0k,i} \\ \Gamma^{i(1)}_{kl} &=& 
\frac{1}{2}(\gamma^{(1)}_{ik,l} + \gamma^{(1)}_{il,k}- \gamma^{(1)}_{ll,i}.
\eeqa
Equation (5.4) shows that the field equations involving $R^{(1)}_{\mu\nu}$ 
are sourceless. The spatial components give 
\beq \gamma^{(1)}_{il,kl}+ \gamma^{(1)}_{kl,il}- \gamma^{(1)}_{ik,ll}
- \gamma^{(1)}_{ll,ik} = 0.\eeq
A lemma by Rendall \cite{rendall1} ensures that any solution to (4.5)
must have the form $\gamma^{(1)}_{ik} = \xi_{i,k} + \xi_{k,i}$. 
Using the gauge transformation (6.2) with $f^k= \xi_k $, we may write
\beq \gamma^{(1)}_{ik} = 0. \eeq 
$R^{(1)}_{0k} = 0$ reduces to 
\beq \gamma^{(1)}_{0l,il}-\gamma^{(1)}_{0i,ll} = 0. \eeq 
$\gamma^{(1)}_{0k}$ transforms under gauges as
\beq \bar{\gamma}^{(1)}_{0k} = {\gamma}^{(1)}_{0k} + h^0_{,k}-f^k_{,0}.
\eeq
with $f^k_{,0l}=0$, see (6.4). 
With a suitable gauge function $h^0$ as solution of 
$h^0_{,kk} = - \gamma^{(1)}_{0k,k}$ , the components $\gamma^{(1)}_{0i}$ 
must be harmonic functions. In view of the discussion in the previous section 
we assume $\gamma^{(1)}_{0i} = const$, the spatial constants (which may be 
time functions) can be transformed to zero with a suitable $f^k$. The last 
equation $R^{(1)}_{00} = 0$ can then be written $\gamma^{(1)}_{00,kk} =0$, 
hence $\gamma^{(1)}_{00} =0$ with a similar argument. Thus the fields 
$h_{(1)}^{\mu\nu}$ or $\gamma^{(1)}_{\mu\nu}$ play only a shadow role:  
Correct at least up to the first post-Newtonian order, we could have 
started with an expansion in powers of the reciproke {\it square} of the
velocity of light. 

\noindent
The first non-vanishing post-Newtonian components of the connection are 
$\Gamma^{\rho (2)}_{\mu\nu}$, which are easily calculated:
\beq  \Gamma^{0(2)}_{00} = \Phi_{,0}~, \Gamma^{0(2)}_{0k} = \Phi_{,k}~,
\Gamma^{0(2)}_{ik} = 0 \eeq
\beqa  \Gamma^{i(2)}_{00} &=& h_{(2)}^{ik} \Phi_{,k} + \gamma^{(2)}_{0i,0} 
- \frac{1}{2}\gamma^{(2)}_{00,i} \\
\Gamma^{i(2)}_{0k} &=& \frac{1}{2}(\gamma^{(2)}_{0i,k}-\gamma^{(2)}_{0k,i}
- \gamma^{(2)}_{ik,0}) \\
\Gamma^{i(2)}_{kl} &=& \frac{1}{2}(\gamma^{(2)}_{ik,l} 
+\gamma^{(2)}_{il,k} - \gamma^{(2)}_{kl,i}) \eeqa
Notice $\gamma^{(2)}_{kl}= - h_{(2)}^{kl}$ from (3.8). We start with the 
equations $R^{(2)}_{ik}=  T^{*(2)}_{ik}$, explicity
\beq \gamma^{(2)}_{il,kl}+\gamma^{(2)}_{kl,il}-\gamma^{(2)}_{ik,ll}-
\gamma^{(2)}_{ll,ik} = 2\Phi_{,ik} +8\pi G\rho \delta_{ik}. \eeq
Here we again use Rendall's lemma in a different form: Denote
the left-hand-side of (4.12) by $p_{ik}$. If $p_{i[k,l]}- 
\frac{1}{4}\delta_{i[k}p_{,l]}=0$, where $p=p_{kk}$, then $\gamma^{(2)}_{ik}$
can be written as $U\delta_{ik}+ \xi_{i,k}+ \xi_{k,i}$. The condition for
$p_{ik}$ is equivalent to (2.14), moreover $U=-2\Phi$. The gauge terms in 
$\gamma^{(2)}_{ik}$ can be transformed away (assume $g^{i}= \xi_{i}$ in (6.3)),
so one is left with
\beq \gamma^{(2)}_{ik} = -2\Phi\delta_{ik}.\eeq
The remaining equations for $R^{(2)}_{0k}$ and $R^{(2)}_{00}$
determine the vector potential $\Psi_{k} \equiv \gamma^{(2)}_{0k}$
and the post-Newtonian scalar potential $\Psi \equiv -\gamma^{(2)}_{00}/2$.
One obtains 
\beq \Psi_{k,ll}-\Psi_{l,lk} = 4\Phi_{,k0}+16G\rho v^k. \eeq
\beq \Psi_{,kk}+\Psi_{k,k0}+3 \Phi_{,00} -2\Phi_{,k}\Phi_{,k}
= 8\pi G\rho(v^2+3p/(2\rho)).  \eeq
Writing $2\Phi_{,k}\Phi_{,k} = (\Phi^2)_{,kk}- 2\Phi\Phi_{,kk},$ 
and using (2.14), which is allowed within post-Newtonian terms, we
have
\beq (\Psi-\Phi^2)_{,ll} + (\Psi_{l,l} + 3\Phi_{,0})_{,0} 
=8\pi G\rho(v^2- \Phi +3p/(2\rho)). \label{psi} \eeq
Defining a potential $\chi$ by
\beq \chi_{,kk} = 2\Phi, \eeq
(4.15) may be written  
\beq (\Psi_{k}-\frac{1}{2}\chi_{,0k})_{,ll} - (\Psi_{l,l}+3\Phi_{,0})_{,k} 
= 16\pi G\rho v^k. \label{psik} \eeq
$\Psi$ and $\Psi_{k}$ are subject to the gauges
\beq \bar{\Psi} = \Psi -k^0_{,0},~~\bar{\Psi_{k}} =\Psi_{k} + k^0_{,k}-g^k_{,0}
\eeq
with $g^k_{,l}= 0$, since $g^k$ was already taken to cancel gauge terms in 
$\gamma_{ik}^{(2)}$. One sees at once that (4.15) and (4.16) resp. (4.17)
and (4.19) are gauge invariant.
For a particular gauge
\beq \Psi_{l,l} + 3\Phi_{,0} =0, \eeq
chosen by Chandrasekhar \cite{chandra}, the field equations simplify
to Laplacians and can easily by solved.

To the field equations one has to add the equations of motion for the fluid. 
To Newtonian order they are given by (\ref{mat}),(\ref{mom}), or explicitly 
in Galilean coordinates:
\beq \rho_{,0} +(\rho v^k)_{,k} =0, 
~~v_{,0}^k + \Phi_{,k} +v^{k}_{,l}v^{l} =0. \eeq
For the post-Newtonian terms we start with the general-relativistic 
expresssions
\beqan u^{\mu}\rho_{,\mu}+ (\rho+\eta^{2}p)u^{\mu}_{;\mu} &=& 0, \\
(\rho+ \eta^2p)u^{\mu}_{;\rho}u^{\rho} 
+(g^{\mu\rho}+ u^{\mu}u^{\rho})p_{,\rho}\eta^2 &=&0 \eeqan
and use the expansion of section 3 as well as (5.2). The result is
again invariant with respect to post-Newtonian gauges:
\beqa
\rho_{,0} +(\rho v^{k})_{,k} + \eta^2(p v^{k}_{,k}-3 \Phi_{,0}\rho
-4\Phi_{,k}v^{k}\rho) &=& 0, \\ 
v_{,0}^i + \Phi_{,i} +v^{i}_{,l}v^{l} 
+ \eta^2(\Psi_{,i}+ \Psi_{i,0}
-3\Phi_{,0}v^i + 2\Phi\Phi_{,i}  \nonumber \\
+(\Psi_{i,k}-\Psi_{k,i})v_k -4v^i \Phi_{,k}v^k
+v^2\Phi{,i} + v^{i}(p_{,0}+ v^{k}p_{,k})/\rho)
 &=& 0. \eeqa

The post-Newtonian field equations (4.19),(4.21) as well as the hydrodynamical
equations (4.23,4.24)  are formally identical with the corresponding 
equations derived by Chandrasekhar \cite{chandra}, provided, the gauge 
(4.21) is adopted (note that in \cite{chandra} an internal energy density is 
introduced, which is neglected here).  

This does not necessarily mean that they also coincide with regard to the 
prediction of experiments involving post-Newtonian gravity, since the basic
framework is different. In particular, as far 
as the comparison of the theory with experiments using light rays is 
concerned, the Newton-Cartan theory must be supplemented by the usual 
Lorentz-covariant electrodynamics instead of its corresponding Newtonian 
limit. This may be done by introducing an ether concept 
in a way also suggested by Trautman \cite{trautman1}. 

Thus it appears that presently there is no practical need to employ the 
Newton-Cartan road to post-Newtonian corrections.   

\sect{Appendix: Matter tensor expansion}
\noindent
For the matter tensor a perfect fluid ansatz will be made. For the components
(3.12) one can write with $\kappa=8\pi G\eta^2$
\beq T_{\mu\nu}^* = 8\pi G[(\rho + p\eta^2)u_{\mu}u_{\nu} 
+ \frac{1}{2}g_{\mu\nu}(\rho -p\eta^2)]\eta^{2}. \eeq
The relativistic four velocity $u^\mu = dx^\mu/ds$ is related to the 
Newtonian velocity $v^\mu = dx^\mu/dt$, where $t$ is the Newtonian
time as defined in (3.18), by $u^\mu = v^\mu dt/ds$. $dt/ds$
is found from (3.3) as
\beq \frac{dt}{ds} = \eta(1-\gamma_{\mu\nu}v^\mu v^\nu \eta^2
- \gamma^{(1)}_{\mu\nu}v^\mu v^\nu \eta^3 + o(\eta^4))^{-1/2} \eeq
Expanding in powers of $\eta$, one obtains for the lowest order coefficients 
of $T^{*}_{\mu\nu}$
\beqa T^{*(0)}_{\mu\nu} &=&  4\pi G \rho t_{,\mu}t_{,\nu} \\
      T^{*(1)}_{\mu\nu} &=&  0 \\
      T^{*(2)}_{\mu\nu} &=&  4\pi G t_{,\mu}t_{\nu}(3p 
      +2\rho \gamma_{\rho\sigma}v^{\rho}v^{\sigma}) 
      + 4\pi G\rho\gamma_{\mu\nu} \nonumber \\
      &&  -8\pi G\rho(t_{,\mu}\gamma_{\nu\alpha}v^{\alpha}+
      t_{,\nu}\gamma_{\mu\alpha}v^{\alpha}).  \eeqa

\sect{Appendix: Gauge transformations}
Like fields also coordinate transformations may be expanded in powers of 
$\eta$. We write for the transformed coordinates
\beq \bar{x}^{\mu} = x_{0}^{\mu} + f^\mu \eta + g^\mu \eta^2 + h^\mu \eta^3 
+ k^\mu \eta^4 + o(\eta^5), \eeq
where all functions on the rhs depend on the original coordinates $x^\mu$.
All expansion coefficients in (3.2), (3.3), (3.11) as well as the singular 
components of the connection (3.10) transform as tensors under the zero-order
transformation $\bar{x}^{\mu} = x_{0}^{\mu}$. Assuming 
$\bar{x}^{\mu} = x^{\mu}$ in (6.1) leads to various gauge transformations. 
From the tensor transformation law for $g^{\mu\nu}$ one obtains (we actually 
need terms up to the sixth order in $\eta$)
\beqa \bar{h}_{(1)}^{\mu\nu} &=& h_{(1)}^{\mu\nu} 
+ f^{\mu}_{,\rho}h^{\nu\rho}
+ f^{\nu}_{,\rho}h^{\mu\rho} \\
\bar{h}_{(2)}^{\mu\nu} &=& h_{(2)}^{\mu\nu} 
+ f^{\mu}_{,\rho}h_{(1)}^{\nu\rho}
+ f^{\nu}_{,\rho}h_{(1)}^{\mu\rho} 
+ f^{\mu}_{,\rho}f^{\nu}_{,\sigma}h^{\rho\sigma}
+ g^{\mu}_{,\rho}h^{\nu\rho}
+ g^{\nu}_{,\rho}h^{\mu\rho} 
\\
\bar{h}_{(3)}^{\mu\nu} &=& h_{(3)}^{\mu\nu} 
+ f^{\mu}_{,\rho}h_{(2)}^{\nu\rho}
+ f^{\nu}_{,\rho}h_{(2)}^{\mu\rho}
+ g^{\mu}_{,\rho}h_{(1)}^{\nu\rho}
+ g^{\nu}_{,\rho}h_{(1)}^{\mu\rho} 
+ f^{\mu}_{,\rho}f^{\nu}_{,\sigma}h_{(1)}^{\rho\sigma} 
 \nonumber  \\ &&
+ (f^{\mu}_{,\rho}g^{\nu}_{,\sigma}
+ g^{\mu}_{,\rho}f^{\nu}_{,\sigma})h^{\rho\sigma} 
+ h^{\mu}_{,\rho}h^{\nu\rho}
+ h^{\nu}_{,\rho}h^{\mu\rho} \\
\bar{h}_{(4)}^{\mu\nu} &=& h_{(4)}^{\mu\nu} 
+ f^{\mu}_{,\rho}h_{(3)}^{\nu\rho}
+ f^{\nu}_{,\rho}h_{(3)}^{\mu\rho}
+ g^{\mu}_{,\rho}h_{(2)}^{\nu\rho}
+ g^{\nu}_{,\rho}h_{(2)}^{\mu\rho}
+ f^{\mu}_{,\rho}f^{\nu}_{,\sigma}h_{(2)}^{\rho\sigma} \nonumber \\ && 
+ h^{\mu}_{,\rho}h_{(1)}^{\nu\rho}
+ h^{\nu}_{,\rho}h_{(1)}^{\mu\rho}
+ (f^{\mu}_{,\rho}g^{\nu}_{,\sigma}
+ g^{\mu}_{,\rho}f^{\nu}_{,\sigma})h_{(1)}^{\rho\sigma} 
+ (f^{\mu}_{,\rho}h^{\nu}_{,\sigma}
+ h^{\mu}_{,\rho}f^{\nu}_{,\sigma})h^{\rho\sigma} \nonumber \\ &&
+ g^{\mu}_{,\rho}g^{\nu}_{,\sigma}h^{\rho\sigma}
+ k^{\mu}_{,\rho}h^{\nu\rho}
+ k^{\nu}_{,\rho}h^{\mu\rho}  \eeqa

\end{document}